\documentclass[single]{cambridge6Atight}
\usepackage{aas_macros}
\usepackage{natbib}

\usepackage{rotating}
\usepackage{floatpag}
\rotfloatpagestyle{empty}
\usepackage{amssymb}
\usepackage{graphicx}

\begin{document}
  

\chapter{Helioseismology: Observations and Space Missions}
\begin{center}
{\normalsize P.~L. Pall\'e$^{1,2}$, T. Appourchaux$^3$, J. Christensen-Dalsgaard$^4$ $\&$ R.~A. Garc\'\i a$^5$ \\
\vspace{0.3cm}
$^1$Instituto de Astrof\'\i sica de Canarias, 38205 La Laguna, Tenerife, Spain\\
$^2$Universidad de La Laguna, Dpto de Astrof\'\i sica, 38206 Tenerife, Spain\\
$^3$Univ. Paris-Sud, Institut d'Astrophysique Spatiale, UMR 8617, CNRS, B\^atiment 121, 91405 Orsay Cedex, France\\
$^4$Stellar Astrophysics Centre, Department of Physics and Astronomy, Aarhus University, Ny Munkegade 120, DK-8000 Aarhus C, Denmark\\
$^5$Laboratoire AIM, CEA/DSM -- CNRS - Univ. Paris Diderot -- IRFU/SAp, Centre de Saclay, 91191 Gif-sur-Yvette Cedex, France
}
\end{center}

\section{Introduction}

The great success of Helioseismology resides in the remarkable progress achieved in the understanding of the structure and dynamics of the solar interior. This success mainly relies on the ability to conceive,  implement, and operate specific instrumentation with enough sensitivity to detect and measure small fluctuations (in velocity and/or intensity) on the solar surface that are well below one meter per second or a few parts per million. Furthermore the limitation of the ground observations imposing the day-night cycle (thus a periodic discontinuity in the observations) was overcome with the deployment of ground-based networks --properly placed at different longitudes all over the Earth-- allowing longer and continuous observations of the Sun and consequently increasing their duty cycles. 

In this chapter, we start by a short historical overview of helioseismology. Then we describe the different techniques used to do helioseismic analyses along with a description of the main instrumental concepts. We in particular focus on the instruments that have been operating long enough to study the solar magnetic activity. Finally, we give a highlight of the main results obtained with such high-duty cycle observations ($>$80\%) lasting over the last few decades.

\section{A historical overview}

The detection of the solar oscillations goes back to more than 50 years ago when \citet{1962ApJ...135..474L}
 discovered the five \-minute oscillations in the solar photosphere. However, it was only with the observations of \citet{1975A&A....44..371D}
that their identity as standing acoustic waves, i.e., 
normal modes of the Sun of high 
spherical-harmonic degree, was established,
confirming previous theoretical inferences by \citet{1970ApJ...162..993U} and \citet{1971ApL.....7..191L},
thus setting the scene for the development of helioseismology.
Coincidentally, a strong inspiration was the announcement in 1975 by the SCLERA
(Santa Catalina Laboratory for Experimental Relativity by Astrometry) 
group of evidence for oscillations in the solar diameter 
\citep[see][]{HILL1976}, of apparently truly
global nature and hence containing information about the entire Sun.
Although these detections were later found not to be of solar origin, 
they contributed substantially to the early investigations of the diagnostic
potential of helioseismology \citep[e.g.,][]{1976Natur.259...89C}.
Low-degree solar five-minute oscillations were announced by \citet{1979Natur.282..591C},
quickly followed by the spectacular observations by \citet{1980Natur.288..541G} from the
geographic South Pole.
With the development of observations of intermediate-degree modes
\citep{1983Natur.302...24D,1985Natur.317..591B,1988ApJ...324.1172L}
the full spectrum of solar oscillations could be identified,
and early helioseismic inverse
analyses yielded the first inferences of solar internal rotation
\citep{DuvDzi1984} and structure \citep{1985Natur.315..378C}.

The early determinations of solar internal rotation gave no evidence for 
a rapidly rotating core with a significant contribution to the oblateness of
the solar gravitational field, thus contributing to the confirmation
through observations of Mercury's orbit of
Einstein's theory of general relativity.
Also, observations by \citet{1989ApJ...336.1092L} of rotational splittings allowed
inversion for the solar internal rotation as a function of latitude and
distance to the centre \citep{1988ESASP.286..149C,1989ApJ...337L..53D}.
The results showed that the surface differential rotation with latitude
persists through the convection zone, while the radiative interior appeared
to rotate essentially as a solid body.
This gives rise to a region of rotational shear at the base of the convection
zone, the so-called {\it tachocline}.

Careful analyses of low-degree oscillations \citep[e.g.][]{1990Natur.347..536E}, which are 
sensitive to the solar core, effectively eliminated modifications
to solar models proposed to account for the low observed solar neutrino 
flux, thus strengthening the case for neutrino oscillations.
This was confirmed by later investigations based on full helioseismic 
inversions \citep{1997PhRvL..78..171B}.

These early results both demonstrated the potential of helioseismology 
and indicated the importance of securing continuous observations over 
extended periods of time.
Radial-velocity observations in integrated light were achieved from the ground by
the International Research on the Interior of the Sun \citep[IRIS,][]{IRIS} and the Birmingham Solar Oscillation Network \citep[BiSON,][still in operation]{1997MNRAS.288..623C}.
Continuous long-term spatially resolved observations were made possible
by the establishment of the Global Oscillation Network Group \citep[GONG,][]{HarHil1996}.
Moreover, the Solar and Heliospheric Observatory (SoHO) satellite \citep{DomFle1995} carry three instruments
for helioseismology: the Global Oscillations at Low Frequency (GOLF) instrument for integrated-light
radial-velocity observations \citep{GabGre1995,GabCha1997},
the Variability of Irradiance and Gravity Oscillations (VIRGO) instrument for irradiance and photometric observations,
also mostly in integrated light \citep{1997SoPh..175..267F},
and the Solar Oscillation Investigation / Michelson Doppler Imager (SOI/MDI) instrument for spatially resolved velocity observations
\citep{1995SoPh..162..129S}.
Of these, GOLF and VIRGO are still in operation,
whereas MDI was switched off after the start of operations
with the Helioseismic and Magnetic Imager (HMI) instrument \citep{HMI} 
on the Solar Dynamics Observatory (SDO) mission \citep{2012SoPh..275....3P}.

%

Although early stability calculations by \citet{1975PASJ...27..581A} indicated that
solar oscillations were unstable in the observed frequency range, later
calculations taking perturbations due to convection into account convincingly
demonstrated their stability \citep{1992MNRAS.255..603B},
implying that the modes are instead excited by the acoustic noise 
generated by the near-sonic speed of near-surface convection.
Early estimates of such stochastic excitation were made by \citet{1977ApJ...212..243G};
they concluded that this mechanism failed to predict the oscillations claimed
by \citet{HILL1976} by a substantial margin.
However, later analyses \citep[e.g.][]{1992MNRAS.255..639B} have shown that stochastic
excitation can account for the observed properties of the oscillations in
the five-minute region, while statistical analysis of the observed 
low-degree modes  was consistent with the stochastic
nature of the mode amplitudes \citep{1997MNRAS.288..623C,FogGar1998}. 

From this evolution has emerged a new and in a true sense more profound branch 
of solar physics. The understanding of the origin of the solar oscillations
has in return inspired the application of similar studies to other stars, 
with {\it asteroseismology}.

\section{Techniques to measure solar oscillations}
\label{sec:instru}
In the Sun, acoustic waves are responsible for the compression and expansion of the gas in the solar atmosphere. An intuitive and straightforward way to detect them is to measure the shifts in the absorption lines --a velocity signal-- in the solar spectrum.  In addition to this effect --because of the radiative damping that takes place as a consequence of the compression-- the brightness increases while it decreases during the expansion. Therefore the solar irradiance is modulated by the effect of the oscillations. For individual acoustic modes, the magnitude of such variations are really small: below $15$ cm/s for velocity shifts and $3 \times 10^{-6}$ for brightness variation ($10^{-3}$ K in temperature). 
Given the very small amplitudes and the required high signal-to-noise ratio,
observation of solar oscillations has required the development of sophisticated
technologies although aided
by the high solar brightness.
Depending on the scientific objectives behind the desired observations, different strategies and instrumentation are possible so the best-suited one could be adopted for each particular case.

\begin{figure}[!htbp]
\center{\includegraphics[width=.99\textwidth]{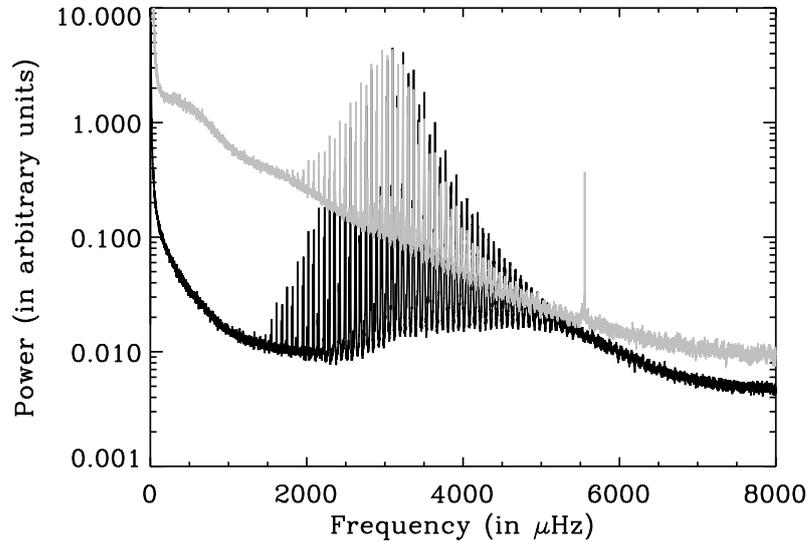}}
\caption[Solar radial velocity vs intensity power spectra]{Power spectra as a function of frequency for the Sun smoothed over 1 $\mu$Hz: solar radial velocity measured by GOLF (black line), and intensity measured by the LOI (gray line).  The GOLF time series is 12.75 year long while the LOI is 16.9 year long.  Power units are arbitrary.  The large spike at 5555 $\mu$Hz is an artefact of the basic cadence of data acquisition system of VIRGO which is 3 minutes.}
\label{golf_loi}
\end{figure}

The main differences between solar radial velocity and intensity measurements are summarised in Fig.~\ref{golf_loi}. This figure shows the comparison between power spectra obtained from Sun-as-a-star velocity observations with the GOLF instrument \citep{GarSTC2005} and power spectra computed with the observations of Luminosity Oscillations Imager \citep[LOI, ][]{1997SoPh..170...27A} of the VIRGO instrument.  We clearly see that the main difference lies in the convective background, i.e., the continuous signal on top of which the discrete spectrum of the acoustic oscillations is seated. Indeed, the granulation is much smaller in solar radial velocity compared to intensity.  As a result, the maximum signal-to-noise ratio of the modes is around 30 in intensity and as high as 300 for solar radial velocity.  By doing radial velocity measurements we are also able to detect more modes at low frequency (typically twice as many) and modes with longer lifetimes up to a year compared to intensity \citep{TouApp1997,2009ApJ...696..653S}.

\subsection{Ground- vs Space-based observations} 

The day/night systematic interruptions generate daily aliases located at harmonics of 11.57 $\mu$Hz (24 hours), making the mode identification difficult when the mode separation is close to that value.  Such interruptions can be overcome by having a network of instruments located all along the Earth as demonstrated in mid 80's \cite{HillNewkirk}.  The typical reduction in the amplitude of the first alias is about $(1-D)^2$, where $D$ is the data fill, i.e. for a 90\% data fill, the alias amplitude is 1\% of the main peak.  The debate gave rise in the 1990's to two types of approaches and thus the subsequent types of projects in helioseismology: ground-based networks (GONG, BiSON, IRIS, TON, etc) and the space-borne mission SoHO with three helioseismic instruments aboard (SOI/MDI, GOLF and VIRGO).  The achieved filling factor $D$ by such projects has been, in some cases, remarkably high ( $>$ 85\%) and maintained for years  and even decades. In addition, avoiding the regularity in the distribution of the gaps clearly reduces the spurious effect of the aliases as proved, first using simulated data \citep{Hill84} and later with real space observations \citep{2014A&A...568A..10G}. 

\subsection{Velocity measurements}
By measuring this physical parameter, a very high signal-to-noise ratio can be
achieved because of the low level of the solar background noise. However, we should take into account the
non-uniform velocity sensitivity across the solar disk arising from the
projection on the line of sight, as well as possible variations in sensitivity induced by the effects of solar
rotation, corresponding to a velocity gradient of roughly
$2 {\rm\, m \, s^{-1} \,arcsec^{-1}}$ \citep{JCDrot}.

\subsubsection{Narrow-Band filters}
\begin{figure}[!htb]
\begin{center}
\includegraphics[width=0.8\linewidth]{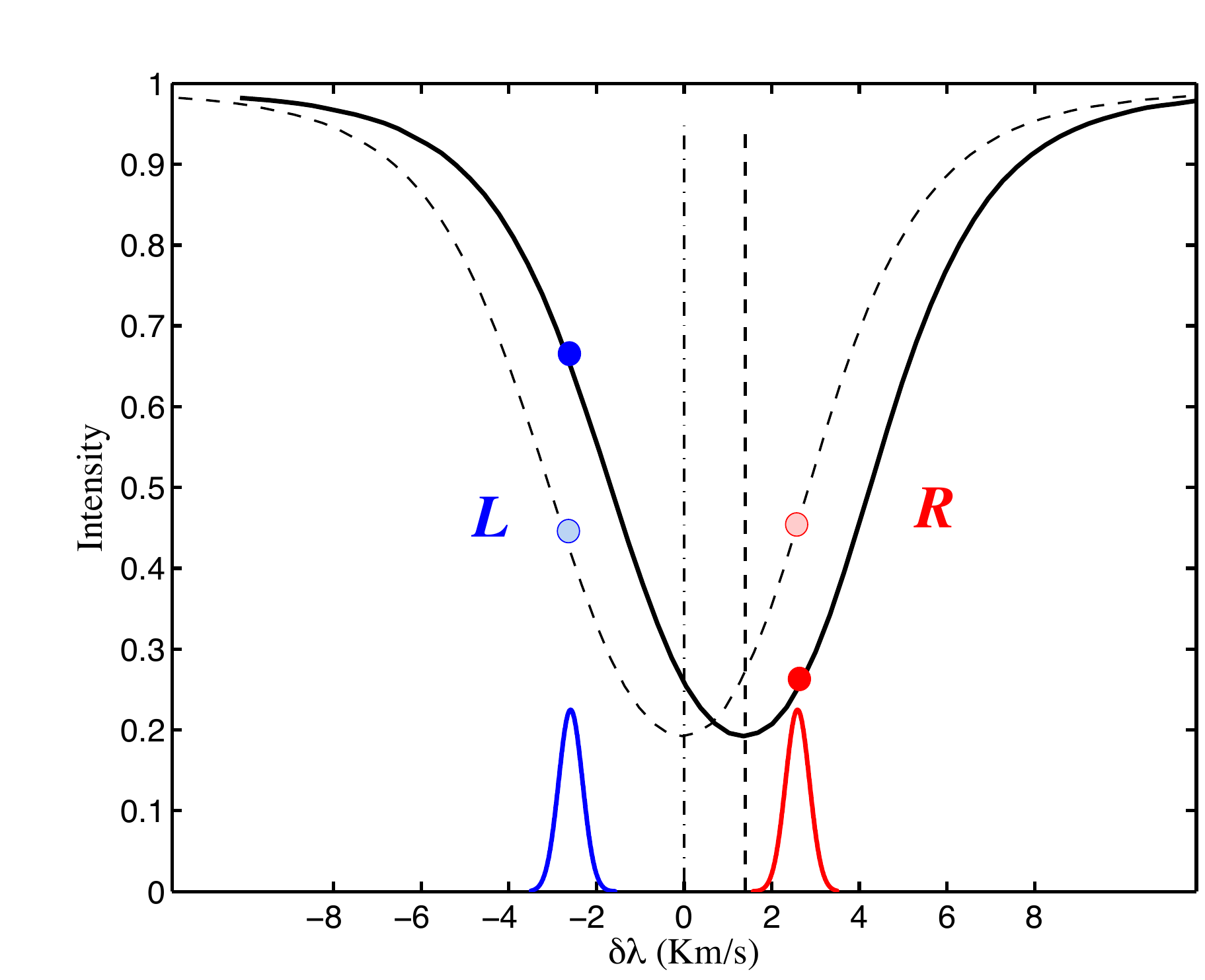}
\end{center}
\caption{\label{fig:line}Principle of operation of the Doppler shift measure by using narrow-band filters. The relative displacement of the solar line (solid) with respect to the laboratory one (dashed-line) can be calculated by measuring at both wings (B and R) and making an appropriate ratio between them.}
\end{figure}

The Doppler velocity shifts are obtained by using narrow-band filters acting on the red ($R$) and/or  blue ($B$) wings
of a selected solar absorption line. As shown in Fig.~\ref{fig:line}, any global shift would increase the transmitted light in one of the wing filters
and decrease the other wing due to the solar line shape. At first order, the difference between the transmitted light in both wings
$( B-R )$ is a measure of the displacement of the solar line relative to its nominal (rest) position.  This approach has
several constraints and disadvantages: some kind of normalization, i.e. ($R+B$), is necessary to reduce transmission changes due to the atmosphere and/or the instrument; variations in the slope of the line profile at the working points of the filter would change the sensitivity; the lack of stability of the bandpasses will produce spurious (fictitious) Doppler velocity signals, etc...

The most representative techniques (and associated instruments) of this group are the following:
\begin{description}
\item[Atomic resonance cell.] This technique has proven to solve (at large extent) the issue of stability and maximize the sensitivity of the velocity measurements, by exploiting the natural stability of resonating atoms.  Briefly, sunlight around a given solar absorption line  (mainly corresponding to alkaline metals like Potassium or Sodium) is properly filtered and circularly polarized. A stable vapour cell of the same chemical element (K or Na), placed in a longitudinal magnetic field, causes the two Zeeman components to sample the line profile. By alternatively measuring the intensities on either side of the line (B and R) the relative position of the solar and laboratory lines may be found and the Doppler velocity shift determined.  First solar spectrophotometers based on this principle were successfully developed and operated in the mid 70Õs by British and French teams \citep{Brookes78, Grec76}. They discovered almost simultaneously the 5-minute solar-oscillations and concluded that they were the normal modes of oscillation of the Sun \citep{1979Natur.282..591C, 1980Natur.288..541G}.  The high sensitivity and robustness of this kind of spectrophotometers resulted in the deployment and operation of  two observational ground-based networks: IRIS \citep{IRIS} and BiSON \citep{1996SoPh..168....1C}, the latter being operational and providing high-quality data since 1976 (more than three solar activity cycles) with an outstanding long-term stability: $\sim$ 1 part in $10^{-3}$ during the last 40 years \citep{GRS}.  In the top panel of Fig.~\ref{fig:ResInst}, the schematics of one of the BiSON nodeÕs instrument, Mark-I, is shown. 

Thanks to the success of this technique and in order to exploit the potential of helioseismic tools, the GOLF instrument (\citealp{GabGre1995}) was chosen as a payload of the ESA/NASA cornerstone mission SoHO. Surrounded by eleven other instruments, GOLF is still providing continuous and useful data since the beginning of operations back in 1996.

The high stability of the Sun-as-a-star observations performed with these instruments allowed us to make a precise determination of the characteristics of low-degree (high spatial scale) modes, in particular those probing the deep layers of the Sun and thus providing structural and dynamical information of these regions.
\begin{figure}[htbp]
\begin{center}
\includegraphics[width=0.95\linewidth]{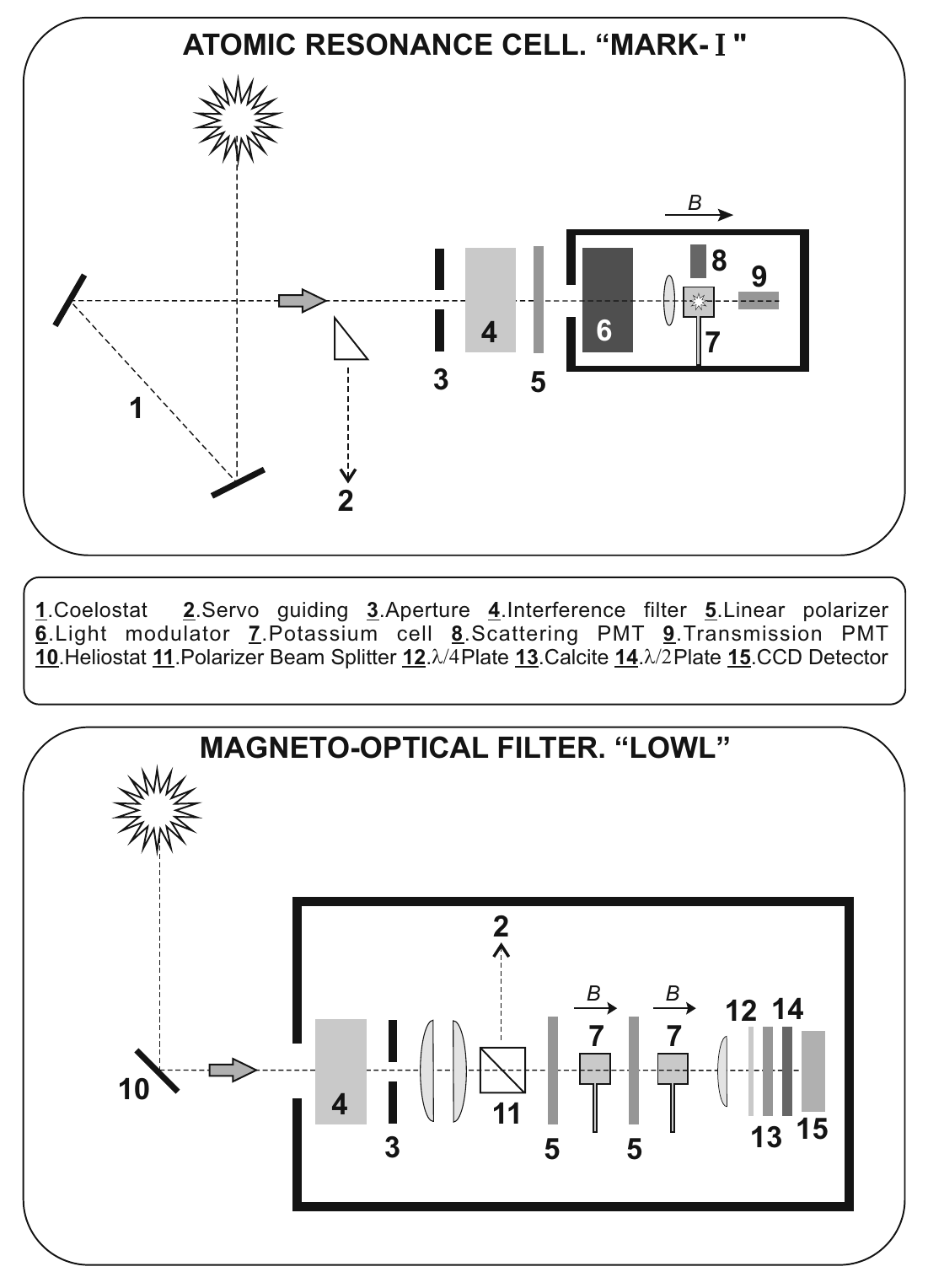}
\end{center}
\caption{\label{fig:ResInst}Schematic representation of the two types of resonant scattering spectrophotometers. The legend box in the middle applies to both schemes. Credits Ramon Castro SMM/IAC.}
\end{figure}

\item[Magneto-Optical Filter (MOF).] This technique is also based on the atomic resonance concept with an imaging capability. By an adequate combination of a double-cell magnet system and polarizing elements (see the bottom panel of Fig.~\ref{fig:ResInst}) global Blue and Red intensity images of the solar surface are obtained.  From these observations, a solar surface velocity map is derived \citep[see for example][]{CaccianiMOF, SteveMOF}.  

This type of instrument provides the so-called medium-degree modes that cannot be observed by Sun-as-a-star observations while being very stable at low frequencies.  

The data and the derived results were really impressive but never crystalized on a robust instrument network capable to achieve high duty cycles for long--term observations (years).  Efforts to deploy networks of MOFs instruments took place by the end of the nineties (HDHN --High Degree Helioseismology Network-- \citealt{RhodesHDHN} and ECHO/LOWL --Experiment for Coordinated Helioseismic Observations- \citealt{SteveECHO}) but only a few years of operations were performed.

A new revised concept is under development (MOTH -- Magneto-Optical filter at Two Heights), which would include various atomic cells with different chemical elements, allowing us to simultaneously sample different heights in the solar atmosphere \citep{MOTH}.

%
\item[ Fourier Tachometer and Michelson Interferometer.] These two technical approaches have many commonalities and represent a powerful way to produce extremely narrow transmission bandpasses on a solar line profile with higher flexibility compared to atomic cells, because they can be conveniently tuned in wavelength. However, they have a much higher technical complexity. The Fourier tachometer Ð-a special purpose of the Fourier transform spectrograph-- provides measurements of the solar line wavelength unaffected of the non-linearity of the line profile and of its variations \citep{BrownFTACH}. Briefly, the incoming solar light is appropriately filtered around the solar absorption line and passes through a single tunable Michelson interferometer. The analysis of its transmission function (with the interferometer properly tuned) provides three independent measurements directly related with the central wavelength, its strength and the brightness of the continuum near the solar absorption line. As a consequence, we can measure Velocity, Modulation and Intensity (V,M,I). This is the Doppler analyser selected for the GONG helioseismology ground-based network \citep{GONG+}: the only ground-based network (with six stations) with imaging capability currently operating and successfully collecting high-quality data since 1996 and achieving nearly continuous observations of the Sun's pulsations.
In order to increase spectral purity of the solar spectral line region, an improved concept was developed where the largest path difference elements in a Lyot birefringent filter are replaced with equivalent elements involving polarizing Michelson interferometers. This combination provides stable and reproducible narrow passbands, well below 100 m\AA $\,$ width. This approach has been used in the MDI instrument \citep{1995SoPh..162..129S} aboard SoHO and, more recently, by its successor, the  HMI instrument  \citep{HMI} aboard the Solar Dynamics Observatory launched in 2010. A schematic layout of HMI instrument is shown in Fig.~\ref{fig:HMIOptic}.  In addition to standard Doppler measurements (with four points of measurement across the line profile) with high spatial resolution on the solar surface, further refinements allowed simultaneous estimates of other quantities such as continuum intensity, line centre depth, and magnetograms. The MDI instrument was switched off after the operations of its successor, HMI that started in 2011. It is providing a nice temporal continuity and extending the capabilities of the SoHO's MDI instrument. It is measuring higher spatial resolution and data cadence as well as high-resolution measurements of the longitudinal and vector magnetic field over the entire visible solar disk 
\end{description}

\begin{figure}[!htbp]
\begin{center}
\includegraphics[width=0.9\linewidth]{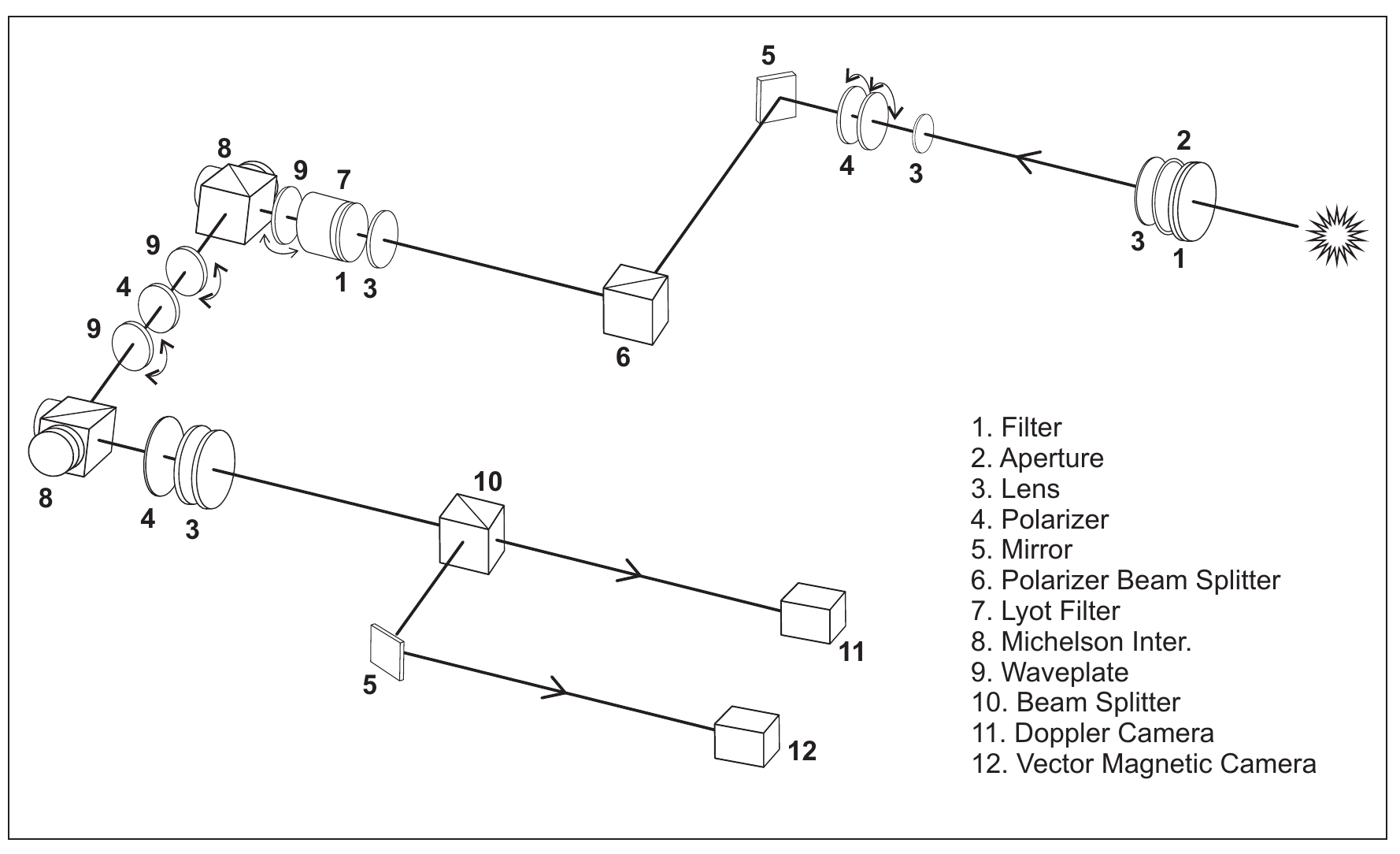}
\end{center}
\caption{\label{fig:HMIOptic} Schematic layout of the HMI Michelson Interferometer  \citep[adapted from][]{HMIdetailed}. Credits Ramon Castro SMM/IAC.}
\end{figure}
%

\subsection{Intensity measurements}
The brightness fluctuations of the solar surface due to the presence of the pressure waves (p modes) is also a physical measurement that carries helioseismic information. Although it may appear rather simple conceptually, the amplitude of many oscillations is close (or even lower) than the intrinsic noise limits (solar background) therefore a precise detection technique is required. However, by measuring the brightness fluctuations instead of line-of-sight velocity, sensitivity to fluctuations is almost uniform over the entire solar disk, including limb fluctuations where line-of-sight velocity measurements are almost insensitive.

\subsubsection{Broad-band intensity.} 
Observations of the continuum intensity of the Sun-as-a-star have been carried out mainly outside the EarthÕs atmosphere in order to avoid the fluctuations of the atmospheric transmission that, ultimately, limits the detection of solar oscillations. They were first measured with radiometers (most famous being the Active Cavity Radiometer -ACRIM type \citep{ACRIM} placed in several spacecrafts for the last decades: SMM, IPHIR, etc. Although the much higher solar background noise in intensity limits the total number of individual modes detected in the spectrum, these observations should help in the detection of solar gravity modes. This is the reason why the SoHO spacecraft included the VIRGO instrument \citep{1995SoPh..162..101F}. It is composed of radiometers, broad-band photometers, and the LOI instrument with a specially designed detector to allow the identification of solar modes up to the degree of 9 (instead of degree 4 for classical disk-integrated measurements).

\subsubsection{Line-core intensity.} The limiting factors in the case of broad-band photometry (small amplitudes of the fluctuations and a large background contribution from convection) can be minimized by observing at a high altitude in the solar atmosphere (core of the spectral line) thus increasing substantially the signal-to-noise level. This kind of observations has been performed mainly in the Ca II K line by using  simple and robust designed instrumentation: a telescope and a suitable designed filter. The pioneer instrument located and operated at South Pole during austral summers \citep{DuvHar1986} demonstrate the potentiality of this technique and of the science behind.  It was followed by a  number of similar single-site instruments \citep{HLH, POI} as well as ground based networks such as the Taiwan Oscillation Network \citep[][]{TON}.

\section{ Summary and future prospects}
\begin{figure}[!htbp]
\begin{center}
\includegraphics[width=1.02\linewidth]{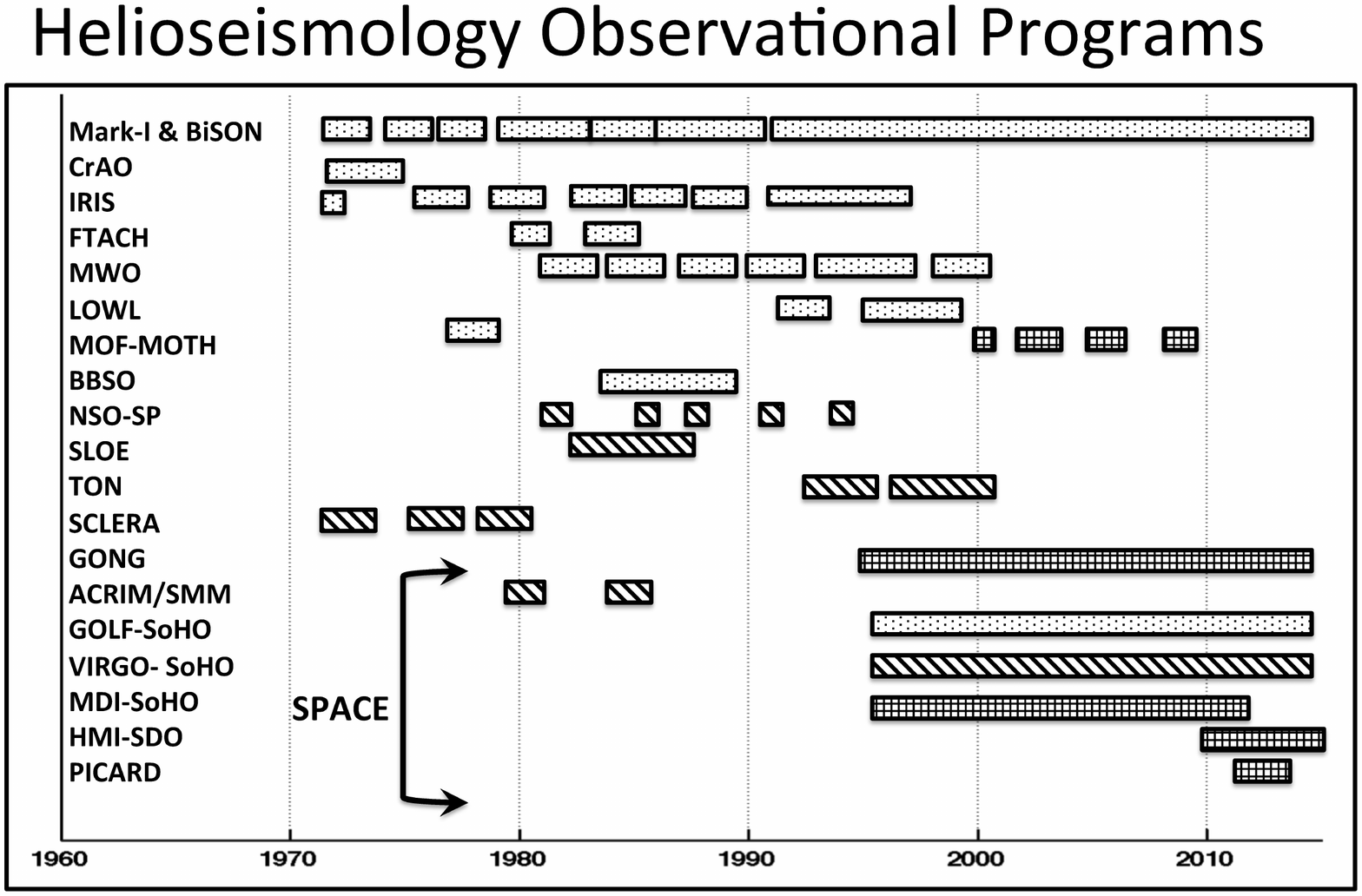}
\end{center}
\caption{\label{fig:FigProjects} List of the main observation initiatives in helioseismology covering the last 50 years. The different shades identify the physical magnitude observed in each project: Doppler velocity (dotted area), Intensity (downward-sloping lines), or both simultaneously (gridded area). Ground- and Space-based programs are clearly identified.}
\end{figure}

During the last 50 years, a large number and variety of instrumentation has been conceived and has been operating. This indicates the wealth of helioseismology that represents the key tool to obtain detailed information on the dynamics and structure of the Sun all the way from the solar atmosphere down to the core. In Fig.~\ref{fig:FigProjects}, recognizing the effort of many instrument's builders and observers, a summary of past and present instrumental initiatives in the field of Helioseismology is shown. Credits Ramon Castro SMM/IAC.

The science objectives for future helioseismic observations are: full disk Doppler velocity, vector magnetic field images, and intensity images, at a variety of wavelengths, high temporal cadence ($<$ 60 s), high duty cycle ($>$90 \%), and at least 25 year lifetime to cover a full solar activity cycle. For the technical requirements such an instrument would require a number of conditions: precise determination of the spectral lines, at least a resolution of 4k by 4k, an aperture of at least 0.5m, adaptive optics, high-speed image post processing, and a high stability.


\bibliographystyle{cambridgeauthordate-rachel}
\bibliography{./BIBLIO}

\end{document}